# Multiple Phases with the Same Quantized Hall Conductance in a Two-Subband System

X. C. Zhang, D. R. Faulhaber, and H. W. Jiang

*Department of Physics and Astronomy, University of California at Los Angeles,
405 Hilgard Avenue, Los Angeles, CA 90095, USA*

Abstract

In a GaAs/AlGaAs two-dimensional electron system with two occupied subbands, the experimentally determined phase diagram in the density-magnetic field plane exhibits rich topological features. Ring-like structures are observed at even integer filling factors, in the phase diagram. Even with the identical quantized Hall resistance values as those given rise by the ordinary integer quantum Hall effect due to the Landau level quantization; the activation energies of these states within the rings are much smaller. These ring structures cannot account for by the simple single particle picture. We argue that ferromagnetic quantum Hall states, due to the interaction of two energy levels with opposite spin and different subband indices, are responsible for these unusual structures.

Quantum Hall physics of two-dimensional electron systems (2DES) is normally studied with semiconductor heterostructures containing a single set of Landau levels[1]. However, presence of a second set interacting of Landau levels can enormously influence the ground state properties of the 2DES. For instance, in the wide single quantum well structure and in the doubled quantum well structure, broken symmetry ferromagnetic phases[2,3] and excitonic superfluid (Bose-Einstein condensate)[4], due to the interaction of the two sets of spatially separated Landau levels, have been realized. The observation of these novel ground states have gained much attention recently. Another semiconductor structure containing two sets of interacting Landau level is a high-density 2DES where both the first and second subband of the confined quantum well are occupied. In this system, due to the additional inter-subband scattering, the mobility of the 2DES is relatively low. On the other hand, since the wave function of the first and second subband are spatially located in the same well, direct spatial coupling between the two sets of Landau levels can potentially give rise to a form of correlation different from the spatially separated cases.

In an earlier study, an experimental phase diagram was mapped out for a single quantum well sample with doubly occupied subbands[5]. The topology of the phase diagram, in the density-magnetic field plane, exhibited "sawtooth-like structures" interpreted as a consequence of Landau level mixing due to disorder broadening. This Landau level mixing was modeled in the framework of a non-interacting particle density of states, which describe the observations adequately.

Here, we explore another double subband single quantum well, a much higher mobility than used in the previous study. The experimental phase diagram obtained in this cleaner system, in contrast to the early observation, shows pronounced "ring-like structures". Unlike for the more disordered sample, we found that these intricate topological structures cannot be accounted for by the single particle picture.

The sample used in this study is a symmetrical modulation-doped single quantum well with a width of 240 Å. Two Si δ-doped layers with doping level $n_d$=10$^{12}$ cm$^{-2}$ are placed on either side of the well. There is a 240 Å spacer between the δ-doped layer and the well on each side. Heavy doping creates a very dense 2DEG, resulting in the filling of two subbands in the well as illustrated in inset (a) of Fig. 1. As determined from the Hall resistance data and Shubnikov-de Haas (SdH) oscillations in the longitudinal resistance, the total density is $n = 8.1 \times 10^{11}$ cm$^{-2}$. The first subband has a density $n_1 = 5.4 \times 10^{11}$ cm$^{-2}$, while the second subband has a density of $n_2 = 2.7 \times 10^{11}$ cm$^{-2}$. The electron mobility at zero gate voltage is about $4.1 \times 10^5$ cm$^2$/V.s, which is extremely high for a 2DEG with two filled subbands. The mobility for each subband is about the same, indicated by absence of noticeable positive magneto-resistance effect near zero field regime. It is worth noting, in the current sample due to higher mobility all the spin splitting are well resolved above 2 Tesla for both subbands at 0.3 K, unlike samples used in a previous study[5]. The samples are patterned into 270×100 μm$^2$ Hall bars using standard lithography techniques. A NiCr gate is evaporated on top of the sample, approximately 3500 Å away from the center of the quantum well. By applying a negative gate voltage on the gate, the carrier density can be varied continuously. A total of three samples from the same wafer were studied, and all have produced remarkably identical results. For consistency we present here the data from only one sample. During the experiment, the sample was placed in the liquid of a top loading $^3$He refrigerator. Standard lock-in techniques with an excitation frequency of about 13 Hz and a current of 100 nA were employed to carry out the magnetoresistance measurements.

As shown in inset (b) of Fig. 1, at zero magnetic field a scan of the gate voltage $V_g$ reveals a 'hump' in the longitudinal resistivity $\rho_{xx}$, absent in one-subband systems. This is a direct evidence of a populated second electric subband well documented in the literature.[6] When $V_g$ is small the second subband is occupied, therefore $\rho_{xx}$ increases monotonically while the 2DEG density is reduced. However, just when the second subband is depleted with $V_g \approx -1.3$ V, the longitudinal resistivity drops suddenly as the system makes the transition from two-subband to single subband occupation, and the inter-subband scattering channel has been suppressed.

Fig. 1 shows the typical traces of longitudinal $\rho_{xx}$ and transverse $\rho_{xy}$ resistivities as a function of magnetic field at $V_g = 0$ at 340 mK. Very pronounced SdH oscillations with an onset of 0.2 T and well developed quantum Hall plateaus were observed. One distinct feature of the two-subband system is existence of more than one frequency component in the oscillations.

Numerous scans were taken by fixing the magnetic field and sweeping $V_g$, producing the gray scale plot of $\rho_{xx}$ as shown in Fig. 2. In the gray-scale map, bright lines represent peaks in $\rho_{xx}$ as shown in Fig. 1, while the dark regions are the minimums, corresponding to the dissipationless quantum Hall states. The most pronounced features in this diagram is the set of ring-like structures centered on $B$ = 4.8 T / $V_g$ = -0.7 V, 3.2 T/-0.8 V, and 2.6T/-1.0 V etc. The centers of these sets of ring-like structures exist on a straight line, which can be extrapolated to a gate voltage value of -1.3 V, coinciding to the depletion of the carrier in the second subband. Another set of rings, which are less distinct, can be seen in more positive $V_g$. The centers of this set fall on another straight line with a larger slope that has the same origin. In the single particle picture, the values (n, B) of these centers correspond to the crossing points of two spin-degenerate Landau levels originated from the two different subbands. So it is not too hard to visually identify the origin of these unusual features being the result of interaction between two sets of Landau levels from two different subbands.

To identify the different quantized states, one can simply read off the quantized values in units of $e^2/h$ in $1/\rho_{xy}$. In Fig. 3, $1/\rho_{xy}$ is plotted in the gray-scale in the $V_g$ -B plane. Fig. 3a focuses on the most pronounced ring structure, followed by the next cleanest one in Fig. 3b. The plots show distinct boundaries between different color regions, corresponding to the labeled quantized values. The ring region in the Fig. 3a has a quantized value of 6 while Fig. 3b has a value of 8. The most intriguing part is the ring being weakly connected to, via almost a point, the other two quantum Hall states with the exact same quantized value. We choose to label these three weakly connected regions as 6A, 6B, and 6C, and 8A, 8B, and 8C.

Since the regions of A, B and C appear to be separable; it is natural to obtain a measurement of their energy scales, associated with the quantum Hall states. In Fig. 4 we present the Arrehenius plots with $log(\rho_{xx})$ as a function of inverse temperature $1/T$. Data is taken from different points as indicated by dots in Fig. 3. Activation energies, $E_a$, at these points can be readily obtained from the slopes. Energy values are 4.8 K, 12.5 K, and 15.4 K for 6A, 6B, and 6C, respectively, and 1.5 K, 8.8 K, and 9.9 K for 8A, 8B, and 8C, respectively. While the energies in the region B, and C are almost identical, the energy in the ring region is significantly smaller by a factor of 3 for the states with $6e^2/h$, and a factor of 5 for the states with $8e^2/h$.

To understand the topological features of the phase diagram, we have performed a simple numerical calculation, analogous to an earlier study[5], to account for the effect of Landau level interaction of the two subbands. In this calculation, we assume the density of states can be modeled as two sequences of Gaussian functions centered around the respective Landau levels. The width of the Gaussian functions is determined from the conductivity of the sample[7]. For each and every maximum in the density of states, the electron density is calculated for a given magnetic field. Assuming the delocalized states correspond to the local maxima in the density of states, we can obtain a theoretical phase diagram in the $n$-$B$ plane. However; we found in contrast to an earlier study in more disordered samples, the non-interacting model can not describe all the topological features, particularly the newly discovered ring-like structures.

We now present a physical picture to speculate the possible origin of the ring structures. In recent years, it has become increasingly clear ferromagnetic ground states are common in quantum systems when two Laudau levels (LLs) simultaneously approach the chemical potential.[8] As illustrated schematically in Fig. 5, applicable to cases of even-number of filled levels, when the two sets of spin-split LL's are far away from each other (regime 1), then both spin states of the first subband are filled like an ordinary, un-polarized quantum Hall state (ex. 6B). When two LL's with opposite spin and different subband indices approach each other, there is now an opportunity for the electrons in the first subband to occupy the up-spin state of the second subband. By spin-flipping to the second subband, the electrons can actually save exchange energy. This phenomenon can be interpreted as a transition from a paramagnetic to ferromagnetic state. The phase transition is expected to take place when the energy spacing between two LL's is roughly equal to the exchange energy (i.e., point a in Fig. 5) corresponding to point a on the left side of the ring-structure. So the quantum Hall states within the ring are a ferromagnetic state (ex. 6A). Likewise, at the higher B side, when the energy spacing becomes larger than the exchange energy (i.e., point b), there is another transition back to conventional un-polarized quantum Hall states (ex. 6C) where two spin states of the upper subband are occupied.

We believe the transitions described here are very similar to those observed in 2D hole systems where a paramagnetic to ferromagnetic phase transition is also observed[9] when two energy levels with opposite spin and different Landau level index approaches each other by increasing the Zeeman energy using a titled magnetic field. Likewise, in a wide quantum well system with two 2D electron layers separated by a soft barrier, a phase transition has been observed[3] when two Landau levels with opposite spins from the symmetrical and anti-symmetrical states are brought close to degeneracy by magnetic field. Similar reports have also been made on a coupled quantum well system[10] and on a parabolic well.[11]

Our explanation is also supported by the following observations. First, if we assume spin-flip occurs when the spacing of the two levels with opposite spins and different subbands is of order of the exchange energy, we can produce in our numerical simulation hexagon-like structures in the locations of the rings with roughly same areas of the rings. We conjecture that the hexagon shape rather than the ring shape in our simulation is because that we have assumed that the exchange energy is a constant rather than a density/field dependent quantity. Second, the activation energies we measured (shown in Fig. 4) are consistent with the experimental values of exchange energy, which is about several to 10 K.[12] However, it is conceivable that the new behavior reflects, instead, the presence of some other interesting collective states, such as a charge ordered[13] or inter-Landau-level skyrmion state[14].

In conclusion, the *n-B* phase diagram of a high mobility GaAs/AlGaAs heterostructure with doubly occupied subbands has revealed a series of ring-like structures. We have shown these structures result from the crossing of two Landau levels with opposite spin and different subband indices. Since these unusual structures cannot be accounted for in single particle picture, we argue ferromagnetic states are possible ground states, analogous to other 2DES systems with similar Landau level crossings.

The authors would like to thank A. MacDonald, S. Kivelson, and S. Chakravarty for helpful discussions, and B. Alavi for technical assistance. This work is supported by NSF under grant DMR-0404445.

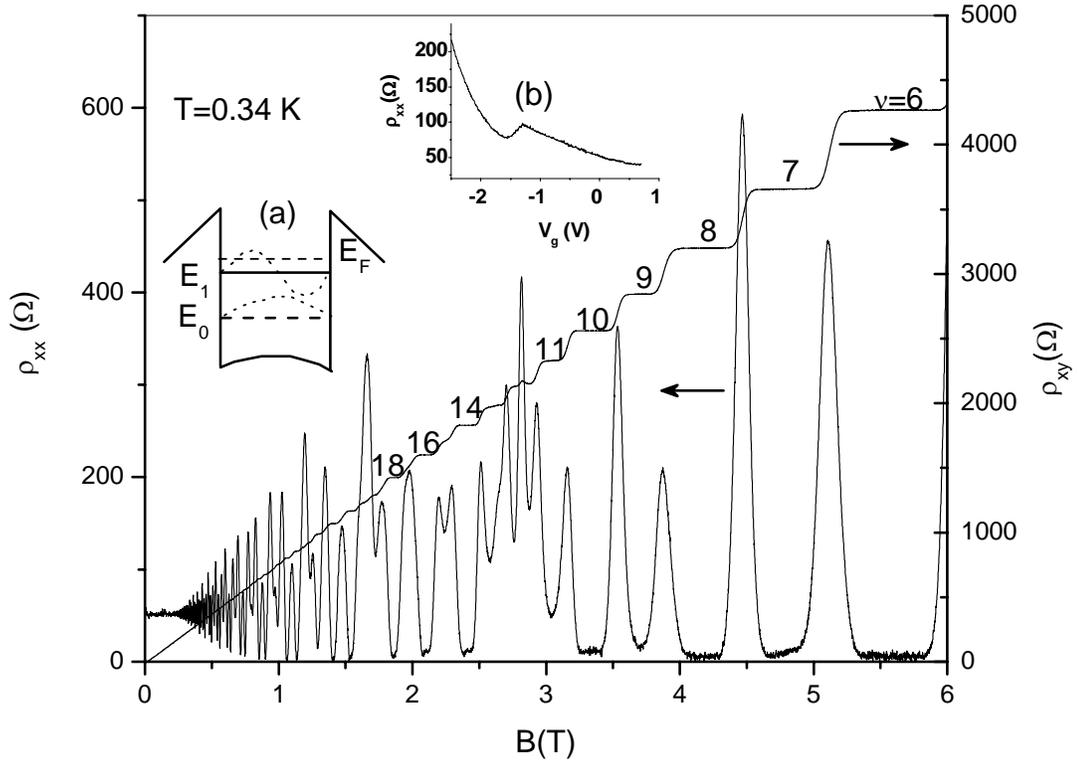

Fig. 1: Typical $\rho_{xx}$ and $\rho_{xy}$ versus B traces at zero gate voltage. Quantum Hall plateaus are labeled by filling factors. Inset (a) schematically shows the band edge profile and the occupied two subbands energy levels and their wave function envelope. In inset (b), the 'hump' in $\rho_{xx}$ versus $V_g$ curve measured at zero magnetic field is indicative of the onset of depletion of the second subband.

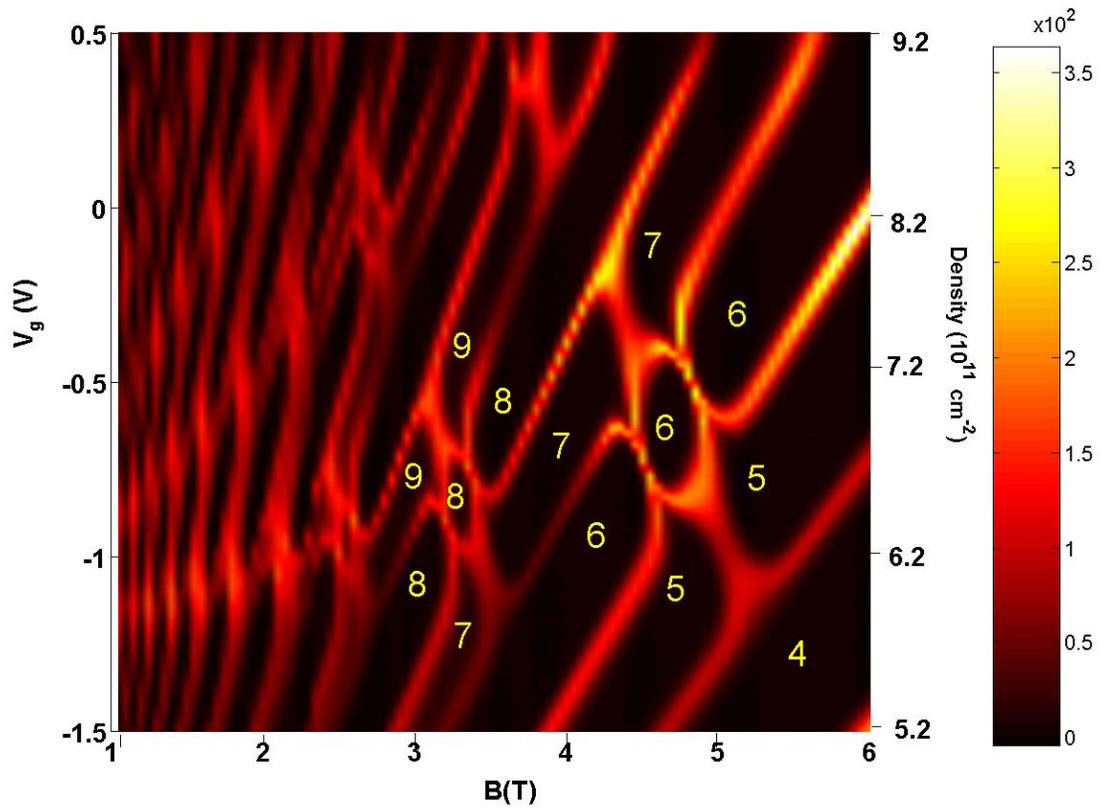

Fig. 2: Gray-scale plot of the longitudinal resistivity $\rho_{xx}$ in the gate-voltage $V_g$ and magnetic field B plane at 0.34 K. Two series of ring structures are seen. The filling factors from $\rho_{xy}$ measurement are labeled.

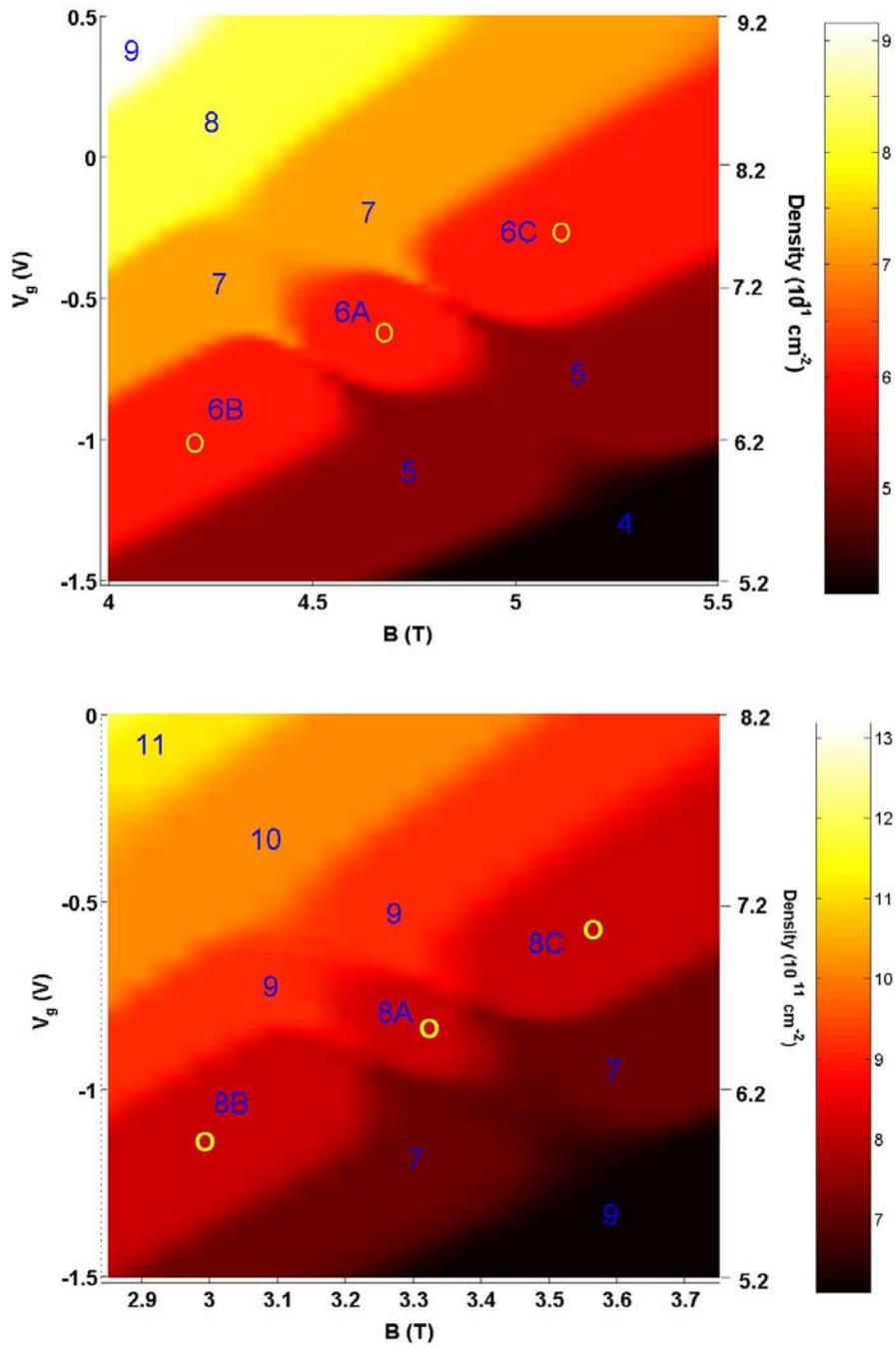

Fig. 3: $1/\rho_{xy}$ is plotted in the gray-scale in the $V_g$ - B plane in the proximity of the two best-resolved ring structures near filling factor 6 and 8. The labels are the quantized values in the unit of e2/h. The center of "O" stands the positions of 6A -0.62V/4.68 T, 6B -1.0 V/4.2 T, 6C 0.25 V/5.16 T, and 8A -0.833 V/3.296 T, 8B -1.125 V/3.03 T, 8C -0.575 V, 3.563 T.

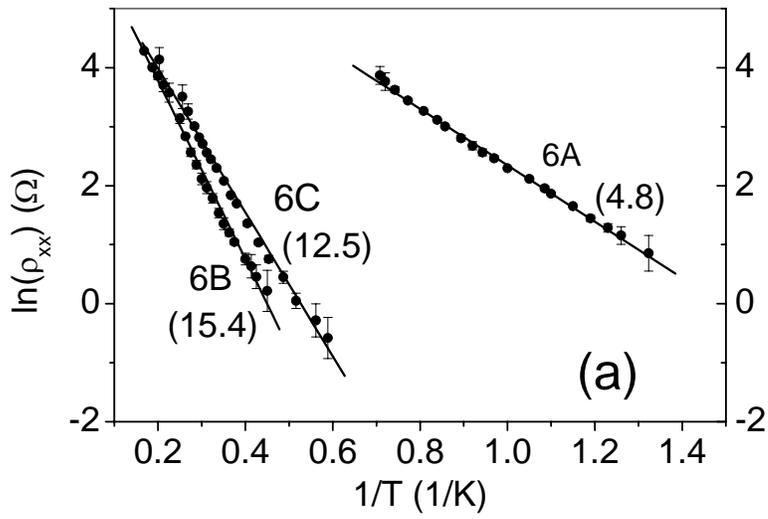

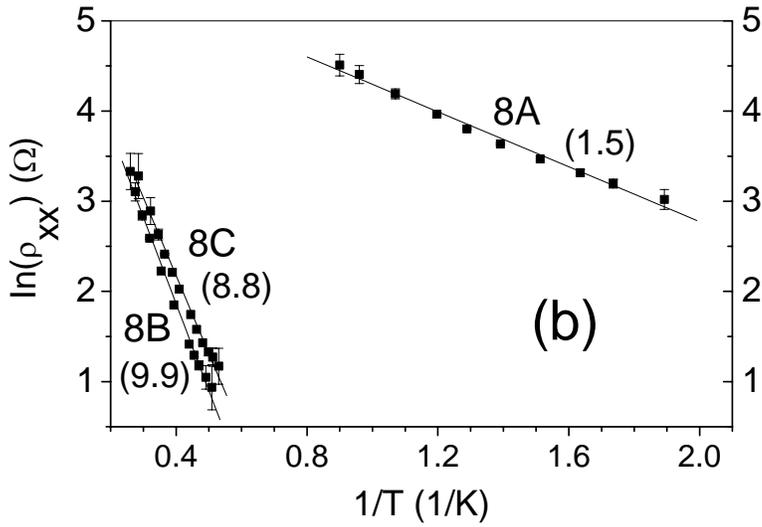

Fig. 4: The log of longitudinal resistivity (filled circles or squares) is plotted against the inverse temperature in the Arrehenius plots for points 6A, 6B, and 6C in Fig. 4(a) as well as 8A, 8B, and 8C in Fig.~6(b). The straight lines are linear regression of experimental data. The numbers in brackets are activation energies in Kelvin.

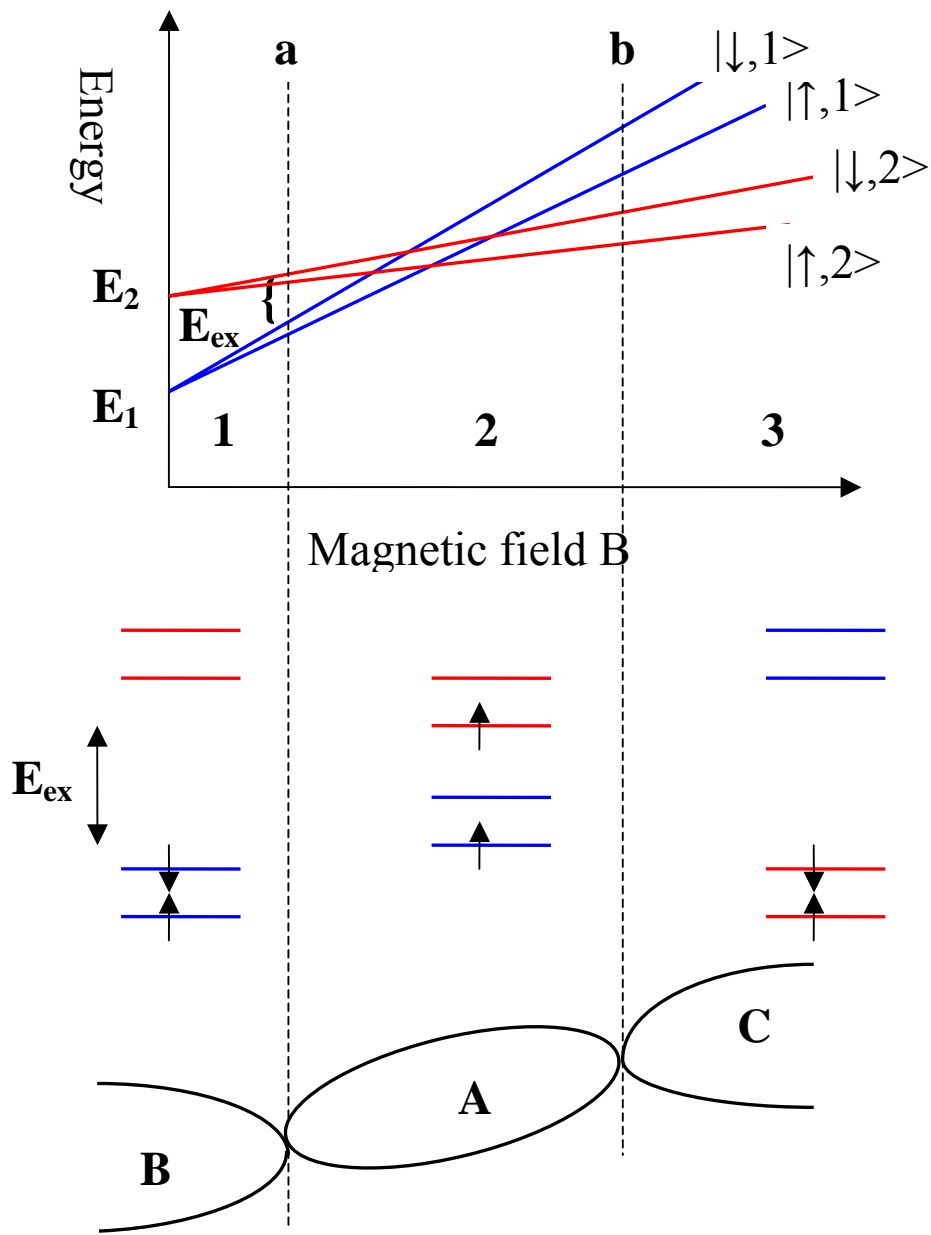

Fig. 5 Energy diagram of two sets of Landau levels that are brought together by the magnetic field. When the spacing of two energy levels with opposite spin and different subband indices becomes smaller then the exchange energy, it is energetically favorable for the electrons to be in a ferromagnetic state (see text).

[1] for a review, see Perspectives in Quantum Hall Effect, edited by S. Das Sarma and A. Pinzuk, (John Wiley, New York, 1997).

[2] T. Jungwirth, S. P. Shukla, L. Smrcka, M. Shayegan, and A. H. MacDonald, Phys. Rev. Lett., 81, 2328 (1998).

[3] V. Piazza, V. Pellegrini, F. Beltram, W. Wegschneider, T. Jungwirth, and A. H. MacDonald, Nature, 402, 638 (1999).

[4] J. P. Eisenstein, Science, 305, 950(2004), and references therein.

[5] X. Y. Lee, H. W. Jiang, and W. J. Schaff, Phys. Rev. Lett., 83, 3701 (1999).

[6] A. R. Hamilton, E. H. Linfeld, M. J. Kelly, D. A. Ritchie, G. A. C. Jones, and M. Pepper, Phys. Rev. B, 51, 17649 (1995), Y. Guldner, J. P. Vieren, M. Voos, F. delahaye, D. Dominquez, J. P. Hirtz, and M. Razeghi, Phys. Rev. B33, 3990 (1986), P. T. Coleridge, Semiconduct. Sci. Technol. 5, 961 (1990), S. E. Schacham, E. J. Haugland, and S. A. Alterovitz, Phys. Rev. B, 45, 13417 (1992).

[7] I. Glozman, C. E. Johnson, and H. W. Jiang, Phys. Rev. B52, R14348 (1995).

[8] T. Jungwirth and A. H. MacDonald, Phys. Rev. B, 63, 035305 (2000).

[9] A. J. Daneshvar, C. J. B. Ford, M. Y. Simmons, A. V. Khaetskii, A. R. Hamilton, M. Pepper, and D. A. Ritchie, Phys. Rev. Lett. 79, 4449 (1997).

[10] A. Sawada et al., Phys. Rev. Lett. 80, 4534 (1998).

[11] G. M. Gusev et al., Phys. Rev. B 67, 155313 (2003).

[12] R. T. Nicholas, R. J. Haug, and K. v. Klitzing, Phys. Rev. B 37, 1294 (1988).

[13] see for example, A. A. Koulakov, M. M. Fogler, and B. I. Shklovskii, Phys. Rev. Lett. 76, 499 (1996); ; M. M. Fogler, A. A. Koulakov, and B. I. Shklovskii, Phys. Rev. B 54, 1853 (1996).

[14] D. Lilliehöök, Phys. Rev. B 62, 7303(2000).